\documentclass{ifacconf}
\usepackage{algorithm}
\usepackage{algpseudocode}
\usepackage{amsmath}
\usepackage{bm}
\usepackage{amsfonts}
\usepackage{graphicx} 
\usepackage{subcaption}
\usepackage{natbib} 
\usepackage{xcolor}

\begin{document}
\begin{frontmatter}

\title{Personalized Building Climate Control with Contextual Preferential Bayesian Optimization\thanksref{footnoteinfo}} 

\thanks[footnoteinfo]{This work was supported by the Swiss National Science Foundation under NCCR Automation, grant agreement {51NF40\_180545} and the Swiss Federal Office of Energy SFOE as part of the SWEET consortium SWICE.}

\author[Shared]{Wenbin Wang} 
\author[Shared]{Jicheng Shi} 
\author[Shared]{Colin N. Jones}

\address[Shared]{Automatic Control Laboratory, École Polytechnique Fédérale de Lausanne, Lausanne, 1015 Switzerland (e-mail: \{wenbin.wang, jicheng.shi, colin.jones\}@epfl.ch)}

\begin{abstract}
Efficient tuning of building climate controllers to optimize occupant utility is essential for ensuring overall comfort and satisfaction. However, this is a challenging task since the latent utility are difficult to measure directly. Time-varying contextual factors, such as outdoor temperature, further complicate the problem. To address these challenges, we propose a contextual preferential Bayesian optimization algorithm that leverages binary preference feedback together with contextual information to enable efficient real-time controller tuning. We validate the approach by tuning an economic MPC controller on BOPTEST, a high-fidelity building simulation platform. Over a two-month simulation period, our method outperforms the baseline controller and achieves an improvement of up to 23\% in utility. Moreover, for different occupant types, we demonstrate that the algorithm automatically adapts to individual preferences, enabling personalized controller tuning.
\end{abstract}


\end{frontmatter}

\section{Introduction}
\label{section 1}
Optimizing occupant utility is one of the main objectives in modern building operation. Buildings account for a substantial portion of global energy consumption \citep{taskgroup2015building}, and numerous studies have demonstrated that efficient operation of heating, ventilation, and air-conditioning (HVAC) systems can significantly reduce energy use \citep{drgovna2020all,5530680,10089206}. However, occupants differ widely in their preferences. For example, some prioritize minimizing energy costs, while others place greater emphasis on thermal comfort. To operate buildings efficiently, control strategies must adapt to these personalized preferences.

Despite this need, most existing controllers are not designed to accommodate individual preferences. Common approaches, including rule-based methods \citep{AGHEMO2013147}, PID control \citep{xu2024data}, and more advanced strategies such as Model Predictive Control (MPC) \citep{10089206}, typically rely on predefined parameters derived from prior knowledge, e.g., a fixed indoor temperature setpoint. Although these methods can effectively reduce energy costs and maintain thermal comfort under nominal conditions, they do not adjust their behavior in response to the preferences of different occupants. For example, reducing indoor temperatures generally lowers energy consumption but increases thermal discomfort \citep{chen2015model}, a trade-off that different occupants value differently. Consequently, a single fixed parameter configuration can be suboptimal across users, which highlights the need for a controller tuning mechanism that can adapt to different types of occupants.

A key challenge lies in quantifying occupant utility, such as thermal comfort, in a reliable manner. Direct reports of utility are typically noisy, biased, and inconsistent \citep{kahneman2013prospect}, which makes them unsuitable for real-time controller tuning. Prior research has shown that people generally provide more reliable preference comparisons than absolute values of their utility \citep{kahneman2013prospect}, suggesting that pairwise preference feedback is a more realistic form of feedback. Nonetheless, many existing studies assume that occupants can directly report their level of discomfort \citep{eichler2018humans}, or rely on voting schemes in which individuals score their thermal sensations \citep{djongyang2010thermal}. Few works have explored controller tuning based on preference feedback, which leaves a methodological gap.

Preferential Bayesian optimization (PBO) \citep{xu2024principled} provides a principled framework for optimizing black-box functions using pairwise comparisons. By assuming that the unknown function lies within a well-structured space, PBO can efficiently identify the optimal set of parameters. However, existing PBO formulations focus on static settings and do not incorporate contextual variables. This limitation is problematic for building applications, where system dynamics are influenced by external factors, such as outdoor temperature and solar irradiation. Seasonal variations also change the building’s thermal response over time \citep{gholamzadehmir2020adaptive}. Furthermore, occupants' behavior, such as opening windows and adjusting blinds, will affect how the controller should behave \citep{afroz2018modeling}. Although the controllers tuned offline exhibit strong performance with the training data, their online performance cannot be guaranteed as the system evolves, which often leads to suboptimal operation.

To address this gap, we propose a contextual Preferential Bayesian Optimization (contextual PBO) algorithm for real-time tuning of building climate controllers. Our method incorporates time-varying contextual information and iteratively recommends controller parameters that optimize occupant utility. We evaluate the approach by tuning an economic MPC controller in real time on BOPTEST, a high-fidelity building simulation platform \citep{blum2021building}. On each day, the occupant provides binary preference feedback based on the latent utility after experiencing different closed-loop trajectories. Then the algorithm proposes the parameters for the next day such that the latent utility can be continually optimized. The results after two months of simulations show that incorporating context enables the algorithm to outperform both the baseline controller and its static PBO counterpart. With the proposed method, we improve the overall utility by up to 23\%. Moreover, we show that the proposed method automatically adapts to two different types of occupants, enabling personalized controller tuning.

The remainder of this paper is organized as follows. Section \ref{section 2} defines the controller tuning problem. Section \ref{section 3} introduces the contextual PBO algorithm. Section \ref{section: 4} describes the simulation setup. Section \ref{section 5} provides the simulation results, followed by conclusions in Section \ref{section 6}.
\section{Problem Setting}
\label{section 2}

\begin{figure}[t]
    \centering
    \includegraphics[width=\linewidth]
    {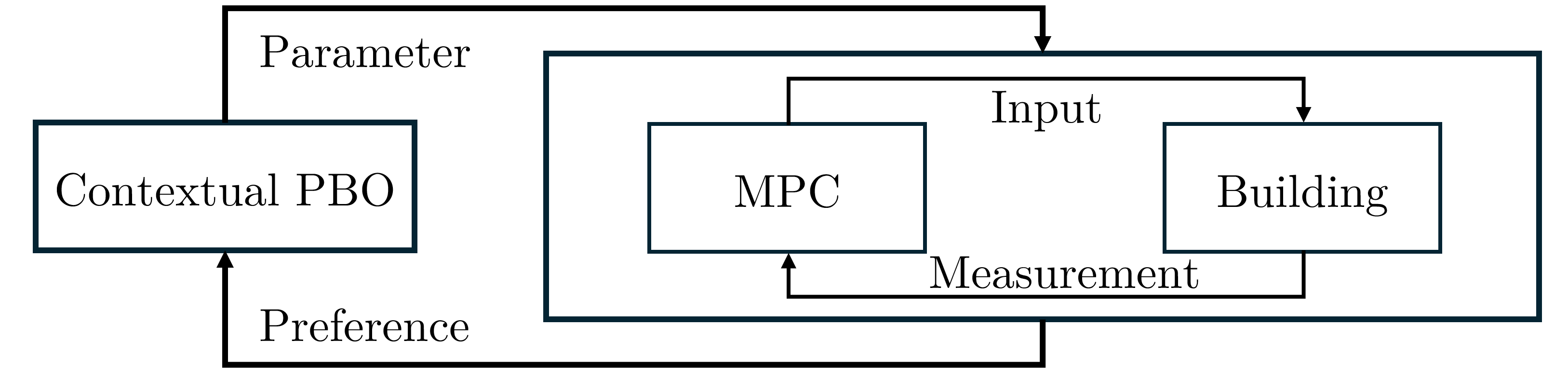}
    \caption{Problem setting for the MPC tuning task.}
    \label{fig: problem setting}
\end{figure}

We aim to tune an MPC controller such that the latent utility is optimized. For each day, the MPC regulates the indoor temperature with a fixed tuning parameter~$\theta$. It is formulated as
\begin{align}
    \min_{\bm{x},\,\bm{u}} \quad 
        & \sum_{k = 0}^{N-1} \ell_k(x_k, u_k, \theta) + \ell_N(x_N,\theta),
        \label{eq:mpc_ocp} \\
    \text{s.t.} \quad 
        & x_{k+1} = f(x_k, u_k, \theta), \forall k \in [N-1] \\
        & \bm{x} \in \mathcal{X}_{\theta}, \nonumber \\
        & \bm{u} \in \mathcal{U}_{\theta}, \nonumber
\end{align}
where the stacked state and input trajectories over the horizon are defined as $\bm{x} = [x_0^\top,\ldots,x_N^\top]^\top$ and 
$\bm{u} = [u_0^\top,\ldots,u_{N-1}^\top]^\top$, respectively. The function $\ell_k$ denotes the stage cost, and 
$\mathcal{X}_\theta$ and $\mathcal{U}_\theta$ represent the state and input constraint sets, respectively. We employ an MPC controller in this work for its ability to explicitly handle constraints and naturally support multi-objective optimization. Nevertheless, the proposed tuning approach is applicable to other types of controllers as well.

The tuning parameter $\theta \in \mathbb{R}^{n_\theta}$ can influence various components of the MPC formulation, including the cost function, the system dynamics, and the constraints. To reduce the problem dimension, it is common to select only the parameters that have the most significant effect on controller performance, such as the lower indoor temperature bound or the weights in the stage cost.

We now formulate the tuning problem as an optimization problem with respect to $\theta$. Let $z_t$ denote the environmental context on day $t$. For each day, we consider
\begin{equation}
    \max_{\theta \in \Theta} \quad J(\theta, z_t),
\end{equation}
where $\Theta$ is the feasible set determined by prior knowledge, and 
$J(\theta,z_t)$ represents the latent utility associated with parameter~$\theta$ under context~$z_t$. We assume that the context lies in a bounded set $\mathcal{Z}$. In practice, $J$ can represent, for example, thermal comfort or energy cost. The context $z_t$ captures uncontrollable external factors, such as outdoor temperature and other disturbances.

On day $t$, once the closed-loop trajectory is realized, the occupant provides pairwise preference feedback comparing day $t$ to day $t-1$. This feedback encodes implicit evaluations of $J(\theta_t,z_t)$ and $J(\theta_{t-1},z_{t-1})$, where $\theta_t$, $\theta_{t-1}$ is the parameter on day $t$ and day $t-1$ respectively. With such feedback, the algorithm then selects $\theta_{t+1}$, thereby iteratively improving the utility. The resulting outer-loop tuning architecture is illustrated in Fig. \ref{fig: problem setting}.

To evaluate the performance, we adopt the metric
\[
R_T^{\text{sum}} = \sum_{t = 1}^{T} J(\theta_t, z_t), \qquad R_T^{\text{ave}}= \frac{1}{T}R_T^{\text{sum}}.
\]
Since the ground truth for the context-dependent black-box problem is generally difficult to obtain, these metrics provide a practical way to assess performance in building thermal control applications \citep{xu2024data}.

\section{Contextual Preferential Bayesian Optimization}
\label{section 3}

Given the preference dataset, a confidence region around the latent utility can be constructed using the likelihood-ratio method \citep{xu2024principled}. This confidence set, combined with a suitable acquisition function, enables sampling strategies that explicitly balance exploration and exploitation. As a result, the algorithm can efficiently identify the optimizer over successive iterations.

We incorporate contextual information by modeling the unknown utility function $J(\theta, z)$ with elements in a reproducing kernel Hilbert space (RKHS) $\mathcal{H}_k$. An RKHS is a Hilbert space of continuous functions in which the point evaluation is a continuous functional \citep{gretton2013introduction}. We assume that the model $\tilde{J}$ has a bounded RKHS norm, $\|\tilde{J}(\theta, z)\|_{\mathcal{H}_k} \le B$, and that the kernel $k$ is continuous and uniformly bounded
\[
    k\big((\theta, z), (\theta', z')\big) \le 1, 
    \qquad \forall \theta, \theta' \in \Theta,\; \forall z, z' \in \mathcal{Z}.
\]

Given the utility $J(\theta,z)$, we say that the pair $(\theta,z)$ is preferred over $(\theta',z')$ if $J(\theta,z)\geq J(\theta',z')$. We denote this event by $(\theta,z)\succ (\theta',z')$. The preference feedback is represented by the indicator variable $\bm{1}_{(\theta,z)\succ (\theta',z')}$, defined as
\begin{equation*}\bm{1}_{(\theta,z)\succ (\theta',z')} = 
\begin{cases}
    1, \quad\text{if $J(\theta,z)\geq J(\theta',z')$},\\
    0, \quad\text{if $J(\theta,z)< J(\theta',z')$}.
\end{cases}
\end{equation*}
We assume that this indicator is a Bernoulli random variable 
with probability
\[\mathbb{P}(\bm{1}_{(\theta,z)\succ (\theta',z')} = 1) = \sigma(J(\theta,z)- J(\theta',z')),
\]
where $\sigma(x)=\frac{1}{1+e^{-x}}$ is the logistic sigmoid function.  If $(\theta,z)$ yields a higher utility, then $(\theta,z)$ is preferred with a higher probability.

Given a preference dataset $\{(\theta_{i},z_{i}),(\theta_{i}',z_{i}'),q_{i}\}_{{i} = 1}^{t}$, the log-likelihood $\ell_t$ given a model $\tilde{J}$ is
\[
    \ell_t(\tilde{J}) = \sum_{i=1}^{t} 
    \bigl(
        q_{i} \tilde{J}_{i} 
        + (1-q_{i}) \tilde{J}_{i}'
    \bigr)
    -
    \sum_{i=1}^{t}
    \log\!\bigl(
        e^{\tilde{J}_{i}} + e^{\tilde{J}_{i}'}
    \bigr),
\]
where we compactly write $q_{i} = \mathbf{1}_{(\theta_{i},z_{i})\succ (\theta_{i}',z_{i}')}$, $\tilde{J}_{i} = \tilde{J}(\theta_{i},z_{i})$ and $\tilde{J}_{i}' = \tilde{J}(\theta_{i}',z_{i}')$. Optimizing over $\tilde{J}$, we obtain the maximum-likelihood value and denote it by $\ell_t^{\text{MLE}}$. The confidence set around the unknown utility function is then constructed as
\[
\mathcal{B}_t:=\{\tilde{J}:\ell_t(\tilde{J})\in [\ell_t^{\text{MLE}}-\beta,\ell_t^{\text{MLE}}]\},
\]
i.e., the set of all candidate functions in the RKHS whose log-likelihood is less than $\ell_t^{\text{MLE}}$ by at most $\beta$. The hyperparameter $\beta$ encodes the confidence level and reflects prior belief about the utility function.

Now, we are ready to present the contextual preferential Bayesian optimization algorithm, summarized in Algorithm~\ref{algorithm:cpbo}.

\begin{algorithm}[H]
    \caption{Contextual PBO}
    \label{algorithm:cpbo}
    \begin{algorithmic}[1]
        \Require $\theta_0$, $z_0$
        \For{$t=1,\ldots,T$}
            \State $\theta_t \gets \arg\max_{\theta\in\Theta} \max_{\tilde J\in\mathcal{B}_t}
                   \tilde{J}(\theta,z_t) - \tilde{J}(\theta_{t-1},z_{t-1})$\label{cpbo:line2}
            \State Query preference feedback $q_t$ and calculate $\mathcal{B}_{t+1}$
        \EndFor
    \end{algorithmic}
\end{algorithm}

At the end of day $t\!-\!1$, the occupant provides feedback by comparing the utility experienced on days $t\!-\!1$ and $t\!-\!2$. This comparison is incorporated into the dataset and used to update the confidence set $\mathcal{B}_t$ (recalculate $\ell_t^{\text{MLE}}$). The algorithm then adopts an optimality-under-uncertainty perspective, i.e., it selects the parameter~$\theta_t$ that yields the largest possible improvement compared to day $t\!-\!1$ over all the functions in the confidence set. Since the context $z_t$ is uncontrollable, the optimization problem is constrained to the observed context for day $t$, as shown in Line \ref{cpbo:line2}. By explicitly incorporating context information, the algorithm learns a context-dependent response surface and adapts in real-time when tuning the controller.

To compute both $\ell_t^{\text{MLE}}$ and the parameter $\theta_t$ in Line \ref{cpbo:line2} of Algorithm \ref{algorithm:cpbo}, we need to solve an optimization over an infinite-dimensional Hilbert space $\mathcal{H}_k$. By invoking the representer theorem \citep{scholkopf2001generalized}, this problem can be reformulated as a finite-dimensional problem. In particular, $\ell_{\text{MLE}}$ can be computed via
\begin{align*}
    \ell_t^{\text{MLE}} \!= \!\max_{\bm{J}\in\mathbb{R}^{t\!+\!1}}\quad &\sum_{i=1}^{t} 
    \bigl(
        q_i\tilde{J}_i
        \!+ \!(1\!-\!q_i) \tilde{J}_{i\!-\!1}
    \bigr)
    \!-\!
    \sum_{i=1}^{t}
    \log\!\bigl(
        e^{\tilde{J}_i} \!+ \!e^{\tilde{J}_{i\!-\!1}}
    \bigr),\\
    \text{s.t.} \quad&\bm{J}^{\top}K^{-1}\bm{J}\leq B^2,
\end{align*}
where $\bm{J} = [\tilde{J}_0,\cdots,\tilde{J}_t]^{\top}$ contains the model values at the observed parameter-context pairs, and $K$ is the kernel matrix defined by $K_{i,j} = k((\theta_i,z_i),(\theta_j,z_j))$.

For fixed  $\theta$ and $z$, a similar representer theorem reduction applies to the optimization problem in Line~\ref{cpbo:line2}, yielding the finite-dimensional formulation
\begin{align*}
    \max_{\bm{J}\in\mathbb{R}^{t+1}}\quad &\tilde{J}_t - \tilde{J}_{t-1},\\
    \text{s.t.} \quad& \ell_{t}(\tilde{J}) \geq \ell_t^{\text{MLE}}-\beta,\\
    &\bm{J}^{\top}K(\theta,z)^{-1}\bm{J}\leq B^2,
\end{align*}
where $K(\theta,z)$ denotes the kernel matrix associated with the candidate parameter $\theta$ and context $z$. This problem is convex with respect to $\bm{J}$. With a low-dimensional parameter space, $\theta_t$ can be located effectively using a grid search over $\Theta$ and choose the one with the largest improvement. In higher-dimensional cases, one can optimize $\theta$ and $\bm{J}$ jointly by employing nonlinear optimization methods with multiple random initializations to mitigate local optima.


\section{Simulation setup}
\label{section: 4}
We demonstrate the advantage of the contextual PBO algorithm by tuning an economic MPC controller in BOPTEST, which is a high-fidelity building simulation platform. Our case study considers the building test case ``{\tt singlezone commercial hydronic}'', which involves a single-zone commercial building equipped with a hydronic radiator heating system and an air-handling unit (AHU), starting on Day~35 of the year. Space heating is provided jointly through the radiator and the warm supply air delivered by the AHU. More details about the simulation configurations and the platform can be found in~\citet{blum2021building}. The MPC controller directly manipulates the radiator valve and is re-tuned on a daily basis. 


\subsection{MPC Formulation}

\subsubsection{Building Modeling}
Building thermal dynamics are predominantly governed by the heat transfer processes and can be approximated effectively by linear RC network models \citep{drgovna2020all}. Motivated by this, we model the building with an AutoRegressive model with eXogenous inputs (ARX) model trained on the historical data. The ARX structure provides a simple yet effective input--output representation and has been widely used in building modeling \citep{WANG2019109405,li2025model}. Specifically, we adopt the form
\[
    y_{k+1} 
    = \bm{a}^{\top}\bm{y}_{k-n_{\text{ARX}}:k}
      + \sum_{i=1}^{n_u} \bm{b}_i^{\top}\bm{u}_{i,k-n_{\text{ARX}}:k},
\]
where $\bm{a}, \bm{b}_i\in\mathbb{R}^{n_{\text{ARX}}}, \forall i \in [n_u] $, $n_{\text{ARX}}$ is the model order, and $n_u$ is the number of exogenous inputs. We set $n_{\text{ARX}} = 10$ and $n_u = 3$. Here, $y_{k}$ denotes the indoor temperature and the exogenous inputs are the radiator valve position, outdoor temperature, solar irradiance, i.e., $u_{i,k}$ for $i = 1,2,3$. The notation $\bm{y}_{k-n_{\text{ARX}}:k}$ ($\bm{u}_{i,k-n_{\text{ARX}}:k}$) denotes the historical output (inputs) from $k-n_{\text{ARX}}$ to $k$.

To obtain an informative and sufficiently excited dataset for system identification, we operate the building under a randomized bang–bang controller acting on the radiator valve starting on Day~35 for one week. When the indoor temperature exceeds a prescribed upper bound, the radiator is switched off. When it drops below a lower bound, it is switched on. When it lies in between, the valve position is toggled randomly. This excitation scheme provides adequate variability in the input signals for robust ARX model estimation.
\begin{figure}[t]
    \centering
    \includegraphics[width=\linewidth]
    {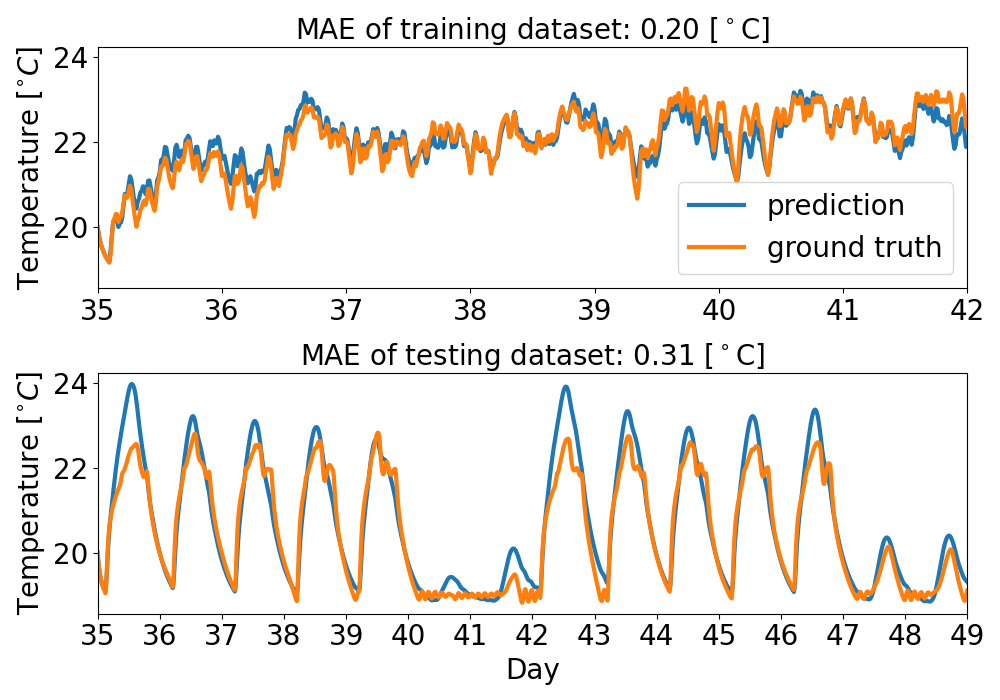}
    \caption{Performance of ARX model.}
    \label{fig: system_id}
\end{figure}

After collecting the training data, a regularized least-squares problem is solved to identify the ARX parameters with all inputs and outputs being normalized. The identified model is evaluated by comparing open-loop predictions against both the training and a separate testing dataset. The testing data are obtained from the BOPTEST baseline controller, which tracks $22 ^\circ\mathrm{C}$ during working hours and $19 ^\circ \text{C}$ during the off-working period.
 
As shown in Fig.~\ref{fig: system_id}, the mean absolute error (MAE) on the training dataset is $0.21^\circ\mathrm{C}$, while the open-loop MAE on the testing dataset, collected over two weeks of operation under the baseline controller, is $0.30^\circ\mathrm{C}$. These results demonstrate that the identified ARX model achieves high predictive accuracy and is well-suited for MPC implementation.

\begin{figure*}[thp!]
    \centering
    \includegraphics[width=\linewidth]{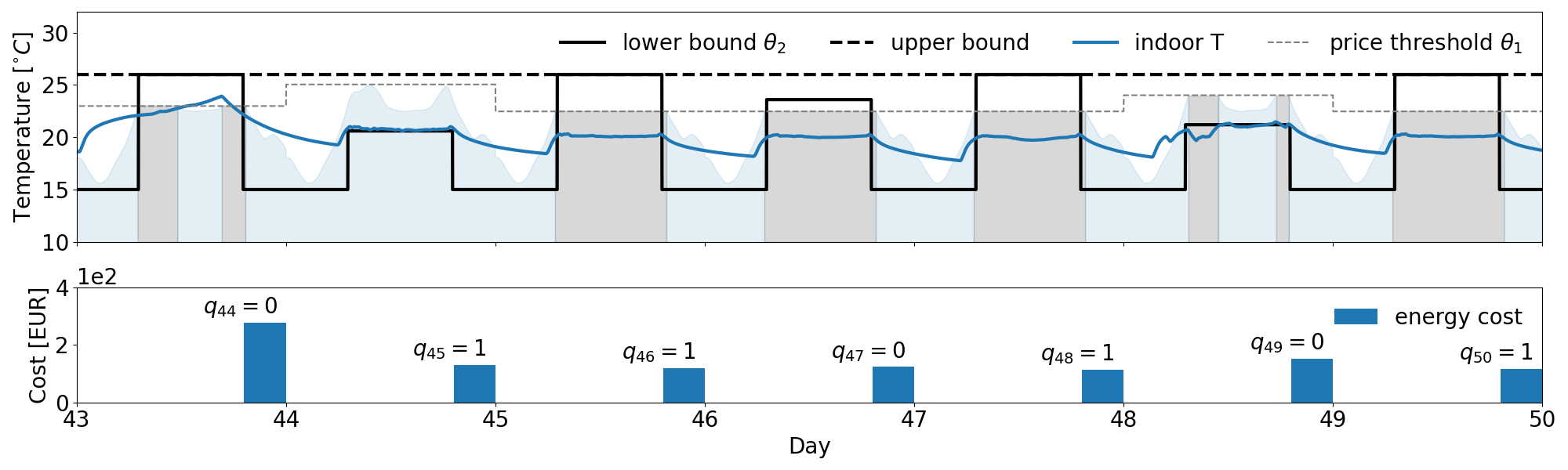}
    \caption{Closed-loop temperature trajectory over one week, starting on Day 43. In the upper subplot, the blue transparent area denotes the daily price pattern, and the grey area denotes the region where the price is higher than the threshold. The lower setpoint is set to $20^\circ\mathrm{C}$ in this case.}
    \label{fig:boptest_result}
\end{figure*}

\subsubsection{Controller Formulation}
We adopt an economic MPC to regulate the indoor temperature, which is formulated as
\begin{align*}
    \min_{u_1,\bm{\epsilon}} \quad 
        & \sum_{k=0}^{N-1} \big( p_k\, u_{1,k} + 1000 \epsilon_k \big) \\
    \text{s.t.} \quad
        & y_{k+1} 
          = \bm{a}^\top \bm{y}_{k-10:k}
          + \sum_{i=1}^{3} \bm{b}_i^\top \bm{u}_{i,k-10:k}, \\
        & y_{k+1} \le 26 + \epsilon_k, \\
        & y_{k+1} \ge \theta_2 - \epsilon_k, \quad\text{if during daytime $p_{k+1}\leq \theta_1$},\\
        & y_{k+1} \ge 20 - \epsilon_k, \quad\text{if during daytime $p_{k+1}> \theta_1$},\\
        & y_{k+1} \ge 15 - \epsilon_k, \quad\text{if during night time},\\
        & u_{1,k} \in [0,1], \epsilon_k \geq 0,\forall k \in \{0,\ldots,N-1\},
\end{align*}

where $p_k$ denotes the heating price and $u_{i,k}$ is the $i$-th input at step~$k$. We read the predicted heating price, outdoor temperature, and solar irradiance directly from BOPTEST and select the price to be a ``highly dynamic'' pattern. Here, $u_{1,k}=1$ represents a fully opened radiator valve and $u_{1,k}=0$ a fully closed valve. The vectors $\bm{y}_{k-10:k}$ and $\bm{u}_{i,k-10:k}$ contain the most recent 10 measurements of the output and of input $i$, respectively, as required by the ARX model. Since the building exhibits relatively slow thermal dynamics, we sample measurements every $900$ seconds and set the MPC horizon to $N  = 64$, corresponding to $16$ hours. A long prediction mitigates myopic decisions and allows the controller to respond proactively to anticipated disturbances and price fluctuations.

The upper and lower bounds on indoor temperature are softened via slack variables to ensure feasibility. The stage cost combines energy cost and slack penalty, with a weight of $1000$ to strongly discourage constraint violations. The upper setpoint is fixed at $26^\circ\mathrm{C}$, following the default BOPTEST configuration, while the daytime lower setpoint is parameterized by~$\theta_2$. To account for varying occupant responses to price fluctuations, we introduce a tuning parameter $\theta_1$, which acts as a price threshold. During daytime, if the heating price $p_k$ is below $\theta_1$, the controller applies the recommended lower setpoint $\theta_2$. Otherwise, a more conservative lower bound of $20^\circ \mathrm{C}$ is used.

\subsection{Occupant Model and Tuning Parameter}
To generate preference feedback from the closed-loop trajectories, we consider two types of occupants. The first type focuses purely on energy cost, with utility defined as
\begin{equation*}
    J(\theta_t, z_t) = -c_t,
\end{equation*}
where $c_t$ denotes the true energy cost. This cost depends on the closed-loop trajectory, which is influenced by the controller parameter $\theta$ as well as the daily context $z$. The energy cost is computed as
\[
    c_t = \int_{\tau} Q(\tau)\, p(\tau)\, d\tau,
\]
where $Q(\tau)$ is the total heating power and $p(\tau)$ is the corresponding heating price at time~$\tau$.

The second type of occupant focuses solely on thermal comfort, with utility defined as
\begin{equation*}
    J(\theta_t, z_t) = -d_t,
\end{equation*}
where $d_t$ is the total thermal discomfort, which consists of two parts. The first part is modeled via the well-known Predicted Mean Vote (PMV) model~\citep{19722700268}. 
Specifically, we use the Predicted Percentage Dissatisfied (PPD) indicator, which returns a value between $0$ and $100$ given an indoor temperature, representing the thermal dissatisfaction. This is defined as
\[
    \frac{1}{\int_{\tau \in \text{daytime}} d\tau}
      \int_{\tau \in \text{daytime}} \mathrm{PPD}\big(y(\tau)\big)\, d\tau,
\]
where $y(\tau)$ is the realized indoor temperature. The integration is restricted to daytime hours when the occupant is present. This normalization yields the average dissatisfaction level over the day. All parameters in the PMV model are kept at standard default values, except for indoor temperature, which is taken from the closed-loop trajectories. 

The second component accounts explicitly for the influence of context. People tend to tolerate lower indoor temperatures when the outdoor temperature is low, for example, by wearing heavier clothing \citep{du2019space}. We model this adaptive behavior using $100(T_{\text{ave.}}-(T_{\text{env.}} + 20))^2$, where $T_{\text{ave.}}$ denotes the daytime average indoor temperature and  $T_{\text{env.}}$ denotes the average outdoor temperature. The motivation is that the preferred indoor temperature decreases when the environment is colder, and the above expression penalizes deviations from this adaptive comfort relation. The total discomfort $d_t$ is the sum of these two terms.

In our real-time MPC tuning framework, the controller parameters are updated daily. We tune the parameters $\theta_1$ and $\theta_2$ jointly, with $\Theta = [0.0889~\text{EUR},\,0.1019~\text{EUR}] \times [20^\circ\mathrm{C},\, 26^\circ\mathrm{C}]$. This ensures that the controller does not drive the indoor temperature to extreme values. The price interval is taken directly from BOPTEST. The context variable $z_t$ is defined as the daily average outdoor temperature $T_{\text{env.}}$, which varies within the range $[-10^\circ\mathrm{C},\, 10^\circ\mathrm{C}]$.

\section{Results}
\label{section 5}
We evaluated the different types of occupants and compared the performance of our method against both the baseline controller from BOPTEST and the static PBO algorithm \citep{xu2024principled}.

\subsection{Energy-Focused Occupant}
In Fig.~\ref{fig:boptest_result}, we first consider an energy-focused occupant. The upper plot shows the indoor temperature trajectories over one week during the online tuning process. The black dotted line denotes the upper setpoint, the black solid line denotes the lower setpoint, and the realized indoor temperature is plotted in blue. The price threshold is represented by the thin dotted line, while the blue shaded area represents the daily price pattern. Each day, the MPC controller is tuned with a different lower bound and price threshold selected by the contextual PBO. When the price is higher than the threshold, the lower setpoint is set to $20^\circ\mathrm{C}$, represented by the gray shaded area.

The lower panel displays the daily energy cost.  During each day, the occupant reports the preference based on the closed-loop cost. For instance, at the end of Day~45, the occupant reports `prefer today’ (i.e., $q_{45} = 1$), since the energy cost on Day~45 is lower than that of Day 44. The higher cost on Day~44 occurs because the algorithm explored a larger lower bound, which required more heating energy. This can also be observed in the elevated closed-loop temperature trajectory shown in the upper plot.

\begin{figure}[t]
    \centering
    \includegraphics[width=\linewidth]
    {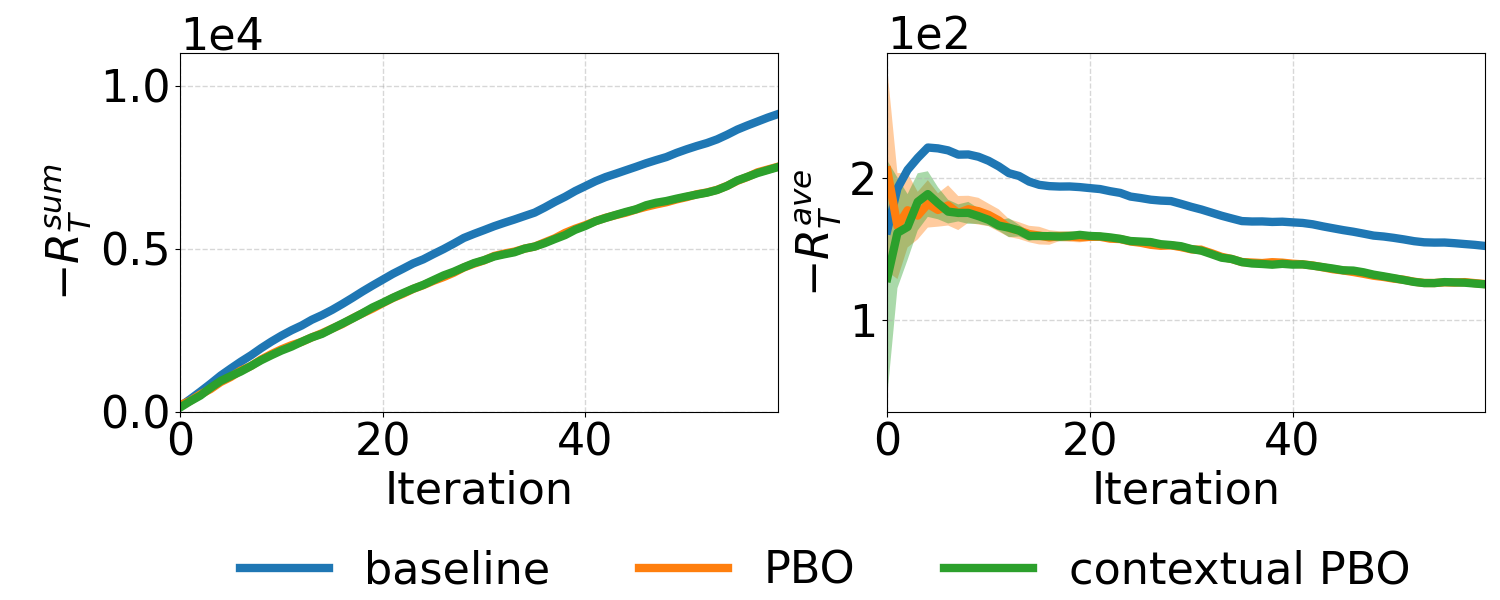}
    \caption{Convergence result for energy-focused occupant.}
    \label{fig: pbo w_10}
\end{figure}

With the same occupant, Fig.~\ref{fig: pbo w_10} shows the cumulative cost and the running-average cost over a two-month simulation period. The green curve corresponds to the contextual PBO algorithm, while the blue and orange curves represent the baseline controller and the static PBO method, respectively. The left plot shows that the baseline controller yields the highest cumulative cost, whereas both PBO and contextual PBO algorithms consistently achieve the lowest cost throughout the simulation. The performance of the two methods is close, and both achieve a cost reduction of up to $18.6\%$ compared to the baseline. This is due to the fact that the optimal strategy is always to choose the lowest price threshold together with a small temperature lower bound. In this regime, contextual PBO provides no disadvantage relative to standard PBO, even though it typically require additional exploration to account for the contextual dimension. The right plot displays the running average cost. All methods show a decreasing trend over time, primarily due to the seasonal transition from winter to spring, during which the energy cost naturally declines as outdoor temperatures rise.

\subsection{Comfort-Focused Occupant}
\begin{figure}[t]
    \centering
    \includegraphics[width=0.8\linewidth]{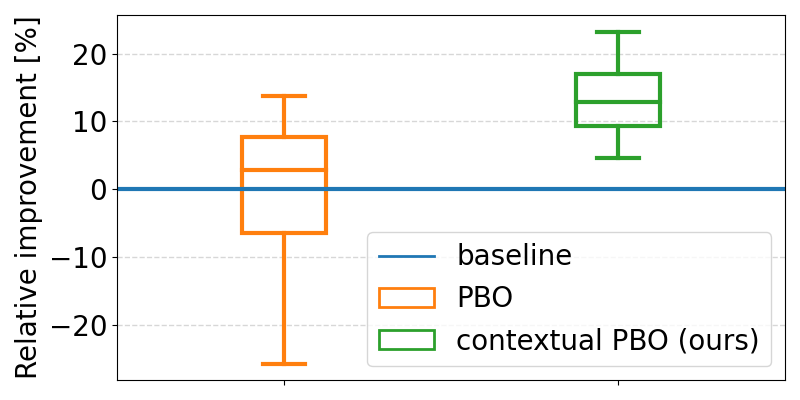}
    \caption{Relative improvement of the cumulative cost $-R_T^{\text{sum}}$ compared to the baseline controller.}
    \label{fig:boxplot}
\end{figure}

In Fig.~\ref{fig:boxplot}, we present the results for a comfort-focused occupant. The relative improvement in thermal discomfort is shown for both static PBO and contextual PBO, evaluated against the baseline controller. Each method is simulated eight times with different random initializations. Since the static PBO does not account for context, it attempts to average across varying environmental conditions, causing some of its recommendations to be suboptimal. This mismatch results in high variability in performance. Consequently, the static PBO has a higher cost than the baseline controller in several simulations. In contrast, contextual PBO updates its learned model dynamically based on the observed context each day, enabling it to recommend more suitable parameters. As a result, contextual PBO achieves utility improvements of up to 23\% compared to the baseline controller.

\begin{figure}[t]
    \centering
    \includegraphics[width=0.8\linewidth]{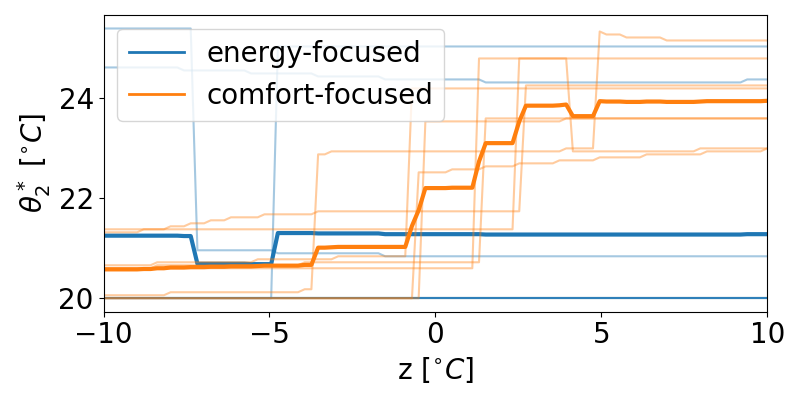}
    \caption{Mapping between context $z$ and the optimal parameter $\theta_2^*$.}
    \label{fig:z_theta}
\end{figure}
To further illustrate that contextual PBO learns a model that adapts to varying environmental conditions, Fig.~\ref{fig:z_theta} shows the optimal lower bound $\theta_2^*$ predicted by the learned model as a function of the average outdoor temperature $z$. The solid orange line corresponds to the comfort-focused occupant, and the solid blue line corresponds to the energy-focused occupant. As discussed earlier, for an energy-focused occupant, the optimal lower bound does not depend on outdoor temperature, which is reflected by the nearly constant blue line.

For the comfort-focused occupant, however, the optimal lower bound increases as the outdoor temperature rises. This aligns with the adaptive comfort model assumption, i.e., when outdoor temperatures are low, occupants tend to tolerate lower indoor temperatures, e.g., by wearing heavier clothing. As the outdoor temperature increases, $\theta_2^*$ also increases. Together, these results demonstrate that the proposed contextual PBO algorithm successfully adapts to different occupant preference profiles and varying environmental conditions, enabling personalized and context-aware controller tuning.

\section{Conclusion}
\label{section 6}
In this work, we addressed the problem of real-time controller tuning with preference feedback by developing a contextual preferential Bayesian optimization algorithm. We validated the proposed method through high-fidelity simulations, in which an economic MPC controller was tuned online over a two-month period. The results demonstrate that the contextual PBO outperforms the baseline controller and achieves an improvement of up to 23\% in utility. Future directions include establishing theoretical guarantees for contextual PBO and deploying the proposed tuning strategy in real building environments.

\bibliography{ifacconf}

\end{document}